\documentstyle[aps,prl,epsf,twocolumn,floats]{revtex}

\begin{document}
\draft
\twocolumn[\csname @twocolumnfalse\endcsname
\widetext

\title{Andreev Reflection and Spin Injection into $s-$
and $d-$wave Superconductors}
\author{Robert L. Merrill and Qimiao Si}
\address{Department of Physics, Rice University, Houston, TX 77251-1892}
\date{\today}
\maketitle

\begin{abstract}

We study the effect of spin injection into $s-$ and $d-$wave 
superconductors, with an emphasis on the interplay between 
boundary and bulk spin transport properties. The quantities of
interest include the amount of non-equilibrium magnetization
($m$), as well as the induced spin-dependent current ($I_s$) and
boundary voltage ($V_s$).
In general, 
the Andreev reflection makes each of the three quantities 
depend on a different combination of the boundary and bulk
contributions. The situation simplifies either for half-metallic
ferromagnets or in the strong barrier limit, where both $V_s$
and $m$ depend solely on the bulk spin transport/relaxation
properties. The implications of our results for the on-going
spin injection experiments in high $T_c$ cuprates are discussed.

\end{abstract}
\pacs{PACS numbers: 71.10.Hf, 73.40. -c, 71.27. +a, 72.15.Gd}
]

\narrowtext
A number of spin-injection experiments have recently
been carried out for the high $T_c$ 
cuprates\cite{Goldman1,Venketesan,Goldman2,Yeh,Hass}.
These experiments are of interest for a variety of reasons. In particular,
they in principle allow us to extract the bulk spin transport properties
of the high $T_c$ cuprates.
In the normal state, such spin transport properties
have been proposed as a probe of spin-charge separation\cite{Si1,Si2}.
In the
superconducting state, they can provide important clues to the nature
of the quasiparticles. In addition, the anisotropy of the spin
transport properties should shed new light on the nature
of the $c-$axis transport. Finally, spin-injection into
superconductors also provides a setting to study the phenomenon of
spatial separation of charge and spin currents\cite{Hershfield,Kivelson}.

For the purpose of extracting the bulk spin transport properties,
it is essential that their contributions to the measured physical
quantities are separated from those of the boundary transport
processes. The boundary of interest here is formed between
the high $T_c$ cuprates and ferromagnetic metals. One 
transport process through such a boundary is the Andreev
reflection\cite{Andreev}, which carries pair-current into
the superconductor. The novel features of the Andreev reflection
involving a ferromagnet have recently been
addressed\cite{Beeneker,Zhu,Valls}.
Another boundary process is the single-particle
transport. While the charge transport involves both processes,
the spin transport proceeds through the single-particle
process only. 
The interplay between the two processes is therefore
expected to play an important role in the spin injection
experiments. The consequences of such an interplay are explored
in this paper.
The $s-$wave case is simpler, which we will address first.
(This part of our analysis is also relevant to the spin injection
experiments in the non-cuprate superconductors\cite{Soulen,Burhman}.)
We will then extend our analysis to the $d-$wave case.

Several physical quantities are of interest in spin injection
experiments. One is the amount of
non-equilibrium magnetization ($m$) injected into the superconductor.
The others include the spin-dependent current ($I_s$) and 
boundary voltage ($V_s$) induced by the injected magnetization.
We will show below that, in general these quantities reflect very
different combinations of the bulk spin transport and interface
transport contributions.

\begin{figure}[t]
\centerline{
\vbox{\epsfxsize=80mm \epsfbox{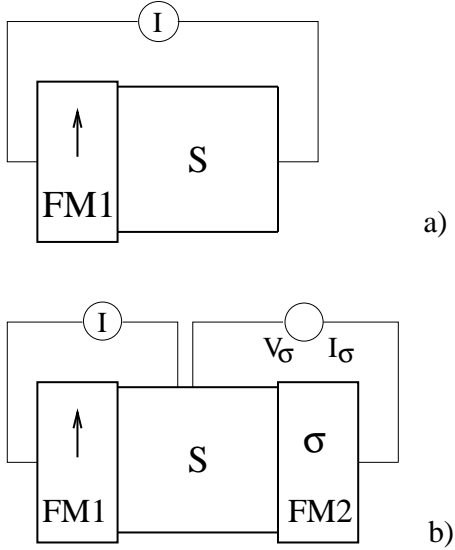}}}
\caption{
a) A spin injection setup involving a ferromagnetic metal (FM1) 
and a superconductor (S). b) A spin-injection-detection setup
involving a superconductor (S) and two ferromagnetic metals
(FM1 and FM2).
}
\label{fig:setup}
\end{figure}

The relevant experimental setups are illustrated in Fig. \ref{fig:setup}.
Fig. \ref{fig:setup}a) is representative of those used in the
on-going spin-injection experiments in the high $T_c$
cuprates\cite{Goldman1,Venketesan,Goldman2,Yeh}.
An injection current ($I$, per unit area) from a ferromagnetic metal (FM1)
to the superconductor (S) is applied, and the critical current
of the superconductor is then measured. The suppression of the critical
current, $\Delta J_c
\equiv (J_c (I=0) - J_c(I))/J_c(I=0)$, is expected to be a measure of
the amount of injected magnetization ($m$). Fig. \ref{fig:setup}b)
illustrates the spin-injection-detection setup of Johnson and
Silsbee\cite{Johnson,Hass}.
Here a superconductor (S) is in contact with two itinerant
ferromagnets, FM1 and FM2.
The magnetization of FM1 is either parallel ($\sigma=\uparrow$)
or antiparallel ($\sigma=\downarrow$) to that of FM2.
For a given $\sigma$, $I_{\sigma}$ is the induced current
across the S-FM2 interface in a closed circuit. Likewise, $V_{\sigma}$
is the induced boundary voltage ($V_{\sigma}$) in an open circuit. 
The spin-dependent current and boundary voltage are
defined as $I_s = I_{\uparrow} - I_{\downarrow}$
and $V_s = V_{\uparrow} - V_{\downarrow}$, respectively.

To highlight the interplay between the boundary and bulk transport processes,
we will make a number of simplifying assumptions.
The superconductor (S) is assumed to be a BCS superconductor,
with either an $s-$wave or a $d_{x^2-y^2}-$wave order parameter.
The ferromagnetic metal will be simply modeled by
an exchange energy $h$\cite{Beeneker,magnon}.
The Hamiltonians for FM1 and FM2 are,
${\cal H}_1 = \sum_{k,\tau} (\epsilon_k - h \tau ) c_{k,\tau}^{\dagger}
c_{k,\tau} $ and ${\cal H} _2 = \sum_{k,\tau} ( \epsilon_k - 
h \tau \sigma) c_{k,\tau}^{\dagger} c_{k,\tau}$, respectively.
The unshifted energy dispersions for both ferromagnets are assumed
to be $\epsilon_k  = {\hbar} ^2 k^2/2m$, as is the normal state
energy dispersion for the superconductor. In addition,
the Fermi wavevectors of the superconductor and the
ferromagnets in the absence of polarization are assumed
to be equal to $k_f$. For FM1, this implies
$k_{f\tau} = k_f t_{\tau}$ and a Fermi 
velocity $v_{f\tau} = v_f t_{\tau}$ where
\begin{eqnarray}
t_{\tau} \equiv \sqrt{1+\tau h / \epsilon_f}
\label{t-sigma}
\end{eqnarray}
For FM2, $k_{f\tau} = k_f t_{\sigma\tau}$ and 
$v_{f\tau} = v_f t_{\sigma\tau}$.

We introduce $m(x)$ to denote
the steady state spin magnetization density
in the superconductor.
Since $m(x)$ is entirely carried by the quasiparticles,
it satisfies
\begin{eqnarray}
-D_s {\partial ^ 2 m / \partial x^2} = -{ {m(x)} / T_1}
\label{kinetic}
\end{eqnarray}
where $T_1$ is the longitudinal spin relaxation time and $D_s$
is the spin diffusion constant.

We now specify the boundary conditions.
For the setup of Fig. 1a), $x=0$ and $x=d$
correspond to the FM1-S interace and the open end of S, 
respectively.
For Fig. 1b), they instead describe
the FM1-S and S-FM2 interfaces, respectively.
At $x=0$, 
\begin{eqnarray}
-D_s \partial m / \partial x |_{x=0} = I_{spin}
\label{bc.spincurrent}
\end{eqnarray}
For the setup of Fig. 1a), and to the
leading order in Fig. 1b) as well,
the boundary condition at $x=d$ is
\begin{eqnarray}
-D_s \partial m / \partial x |_{x=d} = 0
\label{bc.spincurrent2}
\end{eqnarray}

The crucial question then is, what is the spin-current $I_{spin}$
for a fixed injection electrical-current $I$. 
At the FM1-S boundary the electrical current is the 
sum of a pair current ($I_{pair}$), carried through Andreev reflection,
and a single-particle current ($I_{sp}$),
\begin{eqnarray}
I = I_{pair} + I_{sp} 
\label{I-pair-sp}
\end{eqnarray}
The spin current, on the other hand, has no contribution from the
Andreev process given that the Cooper pairs are spin singlets.
We then expect,
\begin{eqnarray}
I_{spin} = \eta \mu_B I_{sp} / e
\label{I-spin}
\end{eqnarray}
where $\eta$ is the single-particle current polarization, which
itself also needs to be determined.
In the following, we first calculate the three currents,
$I_{pair}$, $I_{sp}$, and $I_{spin}$, for a given voltage,
$V$, across the FM1-S barrier. The latter can then be
expressed in terms of $I$, through Eq. (\ref{I-pair-sp}),
leading to an expression for $I_{spin}$ as a function of $I$.

For a fixed 
$V$, the values of the three currents depend on
the nature of the interface.
Here, we model this interface by a delta-function potential,
${\cal H}_{interface} = {\cal V} \delta (x)$.
The Bogoliubov-deGennes equation\cite{BTK}
is,
\begin{eqnarray}
\left( \begin{array}{ll}
        {\cal H}_0 - h \tau ~~~~~~~~~ & \Delta \\
        \Delta ^*~~~~~~~~~ & -({\cal H}_0 + h \tau)
\end{array}
\right)
\left( \begin{array}{ll}
        u_{\tau} \\
        v_{\bar{\tau}}
\end{array}
\right)
=
E
\left( \begin{array}{ll}
        u_{\tau} \\
        v_{\bar{\tau}}
\end{array}
\right)
\label{B-dG}
\end{eqnarray}
where ${\cal H}_0 = p^2/2m -\mu + {\cal H}_{interface}$.
Here, $h$ and $\Delta$ are non-zero only for
$x<0$ and $x>0$, respectively.
Inside FM1, the solution is the sum
of three plane waves, describing the
incident electron of spin $\tau$
and wavevector $\vec{k}_{i\tau}$,
the Andreev-reflected hole of spin $\bar{\tau}$,
wavevector $\vec{k}_{a\tau}$ and amplitude $a_{\tau}$,
and the normally-reflected electron of
spin ${\tau}$, wavevector $\vec{k}_{b\tau}$ and 
amplitude $b_{\tau}$, respectively.
Inside the superconductor, the solution is
the sum of two plane waves, one without
branch-crossing and of wavevector $\vec{k}_{c\tau}$
and amplitude $c_{\tau}$, the other with
branch-crossing and of wavevector $\vec{k}_{d\tau}$
and amplitude $d_{\tau}$. 
The wavevectors are determined by the requirement that,
the components parallel to the interface are equal.
The amplitudes are then calculated by solving Eq. (\ref{B-dG}),
together with the boundary conditions\cite{BTK} appropriate to
this equation.
These amplitudes, in turn, allow us to 
determine the various currents.

In the linear response regime, the pair, single-particle,
and spin currents can be written as
$I_{\lambda}/V = 
\int d E (-\partial f / \partial E) \sum_{\vec{k}\tau}
{\cal F} \delta (\xi_k^{\tau} - E) v_k^x i_{\lambda}^{\tau}
(E, \vec{k})$.
Here, $\xi_k^{\tau} = 
\epsilon_k - \epsilon_f - h \tau$, 
$f(E)$ is the Fermi-Dirac distribution function,
${\cal F}$ is a filtering factor which specifies
the distribution of incident angles. In addition,
\begin{eqnarray}
i_{pair}^{\tau} &&= 2 e^2 A_{\tau} \nonumber\\
i_{sp}^{\tau} &&= e^2(C_{\tau} + D_{\tau}) \nonumber\\
i_{spin}^{\tau} &&= {\mu_B}e (C_{\tau}+D_{\tau})\tau
\label{I-in-terms-of-amplitudes}
\end{eqnarray}
where $C_{\tau} =|c_{\tau}|^2 (u(k_{c\tau})^2-
v(k_{c\tau})^2) v_{f}^x/v_{k\tau}^x$,
$D_{\tau} =|d_{\tau}|^2 (u(k_{d\tau})^2-
v(k_{d\tau})^2) v_{f}^x/v_{k\tau}^x$,
and $A_{\tau} = |a_{\tau}|^2 v_{k\bar{\tau}}^x/v_{k\tau}^x$.
(Here $k\tau$ labels the wavevectors satisfying $\xi_k^{\tau} = E$.)

The potential barrier can be represented in terms of a
dimensionless quantity, $z = m{\cal V}/{\hbar}^2k_f$. 
In the following, we will focus on the two extreme limits,
$z \rightarrow 0$ and $z \rightarrow \infty$. 
In the latter case we expand in terms of $1/z$.
In addition, we will consider only temperatures low 
compared to the superconducting gap $\Delta$. These
cases are sufficient to illustrate the main points
of this paper. More general cases will be discussed elsewhere.

{\em $s-$wave, vanishing barrier:~~}
We address first the case of an isotropic $s-$wave superconductor,
in the limit of a vanishing barrier, $z=0$. For $k_BT \ll \Delta$,
we found,
\begin{eqnarray}
I_{pair}/  V &&= g_ 0P_{s,0} \nonumber\\
{ I_{sp} / V} &&= g_0  S_{s,0} Y_{s,0}(T) \nonumber\\
{ I_{spin} / V} &&= (\mu_B / e) g_0 M_{s,0} Y_{s,0}(T)
\label{I-s-wave-z=0}
\end{eqnarray}
Here $g_0 = N_0 v_f e^2$ where $N_0$ is the density of
states at the unshifted Fermi energy,
$P_{s,0} = 16 \alpha_{pair} 
{{ t_{\uparrow}t_{\downarrow}}/
{(t_{\uparrow}t_{\downarrow} + 1 )^2}}$,
$S_{s,0} = 2 \sqrt{2\pi} \alpha_{sp}
{ {[ t_{\uparrow} (t_{\downarrow}^2 +1)
+
t_{\downarrow} (t_{\uparrow}^2 +1)]}
/ {(t_{\uparrow}t_{\downarrow} + 1 )^2 }}$,
and 
$M_{s,0} = 
2 \sqrt{2\pi} \alpha_{spin} 
{{(t_{\uparrow}- t_{\downarrow})
(1 - t_{\uparrow}t_{\downarrow})}
/ {(t_{\uparrow}t_{\downarrow} + 1 )^2 }}$.
The (polarization-dependent) factors 
$\alpha_{pair}$, $\alpha_{sp}$, and 
$\alpha_{spin}$ are of order unity;
they describe the averaging over the angle of incident electrons,
and are all equal to unity when forward incidence dominates. 

Both $I_{sp}$ and $I_{spin}$ have an exponential temperature
dependence, as specified by the function
$Y_{s,0}(T) = \sqrt{{k_BT}/ \Delta} e ^{-\Delta / k_BT}$.
This reflects the simple physics that the transport of both
the single-particle current and spin current involve 
the quasiparticles of the superconductor and hence only energies
above the superconducting gap. 

We are now in a position to determine the non-equilibrium
magnetization ($m$) and the induced spin-dependent boundary
voltage ($V_s$) and current ($I_s$) as a function of $I$. 
Consider first the general case, when the ferromagnet has a less than 
100\% polarization.
For $k_B T \ll \Delta$, $I_{pair}$ then dominates
the total current in Eq. (\ref{I-pair-sp}).
As a result,
\begin{eqnarray}
I_{spin} / I = (\mu_B / e) (M_{s,0}/P_{s,0}) Y_{s,0}(T)
\label{Ispin-s-wave-z=0-generic}
\end{eqnarray}
Combining Eq. (\ref{Ispin-s-wave-z=0-generic}) with Eqs. (\ref{kinetic},
\ref{bc.spincurrent}, \ref{bc.spincurrent2}), we have,
\begin{eqnarray}
{ m(x=d) / I} = (\mu_B / e) (M_{s,0} /P_{s,0})
Y_{s,0}(T) { T_1  \over {\delta_s \sinh(d/\delta_s)}}
\label{m-s-wave-z=0}
\end{eqnarray}
where $\delta_s = \sqrt{D_s T_1}$ is the spin-diffusion length of
the superconductor. 

This magnetization accumulation leads to a drop across the interface of
an effective magnetic
field\cite{Johnson}, $\Delta H = m /\chi$ where $\chi$ is the uniform
spin susceptibility of the superconductor. 
In the spin-injection-detection setup (Fig. 1b), for a closed circuit,
this effective field drop will induce a current across the S-FM2
interface. Following a procedure similar to
the one leading to Eq. (\ref{I-in-terms-of-amplitudes}),
we found
\begin{eqnarray}
I_{\sigma}^{closed} = 
\sigma g_0 (\mu_B / e) M_{s,0} Y_{s,0}(T) (m /\chi)
\label{I-sigma-closed-s-wave-z=0}
\end{eqnarray}
The induced spin-dependent current, 
$ I_s \equiv I_{\uparrow}^{closed} - I_{\downarrow}^{closed}$,
is then equal to
\begin{eqnarray}
I_{s}/I = 2 g_0 ( \mu_B / e)^2 ( M_{s,0} ^2 / P_{s,0})
[Y_{s,0}(T)]^2 { T_1  \over {\chi \delta_s \sinh(d/\delta_s)}}
\label{I-s-s-wave-z=0}
\end{eqnarray}
Similarly, in the open circuit case, a boundary voltage will 
develop across the S-FM2 interface
to balance the current that would have been induced by the 
effective magnetic field drop.
The total induced current is,
\begin{eqnarray}
I_{\sigma}^{open} = 
\sigma g_0  ( \mu_B /e)
M_{s,0} Y_{s,0}(T) 
( m /\chi) + g_ 0P_{s,0} V_{\sigma} 
\label{I-sigma-closed-s-wave-z=0-open}
\end{eqnarray}
Compared to Eq. (\ref{I-sigma-closed-s-wave-z=0}), 
the additional term is the current induced by the voltage drop;
here only the pair current has been kept 
as it dominates over the corresponding single-particle current.
Setting $I_{\sigma}^{open} = 0$ determines $V_{\sigma}$,
leading to a
spin-dependent boundary voltage, $V_s = V_{\uparrow} - V_{\downarrow}$,
as follows,
\begin{eqnarray}
V_{s}/I = 2 (\mu_B / e) ^2
({ M_{s,0} / P_{s,0}})^2 [ Y_{s,0}(T) ]^2
{T_1  \over {\chi \delta_s \sinh(d/\delta_s)}}
\label{V-s-s-wave-z=0}
\end{eqnarray}

Eqs. (\ref{m-s-wave-z=0},
\ref{I-s-s-wave-z=0},
\ref{V-s-s-wave-z=0}) reveal one key point:
In general each of the three quantities, $m$, $V_s$, and $I_s$,
reflects a different combination of the boundary and bulk
contributions. 

Consider next the special case of a half-metallic ferromagnet.
$I_{pair}$ vanishes in this case, as can be seen from the
expression for $P_{s,0}$ in Eq. (\ref{I-s-wave-z=0});
the Andreev reflection is completely suppressed
in this case\cite{Beeneker}.
The total current in Eq. (\ref{I-pair-sp}) is then entirely
given by $I_{sp}$.
This, combined with the fact that $M_{s,0} = S_{s,0}$
for a half-metallic ferromagnet, leads to a simple relationship
between the spin and charge current, 
\begin{eqnarray}
I_{spin} / I = \mu_B / e
\label{I-spin-s-wave-z=0-hm}
\end{eqnarray}
In addition, 
the  second term in Eq. (\ref{I-sigma-closed-s-wave-z=0-open}) 
is now replaced by the corresponding
single-particle current. The results then become,
\begin{eqnarray}
m(x=d) / I &&= (\mu_B / e)
{ T_1  \over {\delta_s \sinh(d/\delta_s)}}\nonumber\\
I_{s}/I && = 2 g_0 ( \mu_B / e)^2
M_{s,0} Y_{s,0}(T) 
{ T_1  \over {\chi \delta_s \sinh(d/\delta_s)}}\nonumber\\
V_{s}/I && = 2 (\mu_B /e) ^2
{ T_1  \over {\chi \delta_s \sinh(d/\delta_s)}}
\label{results-s-wave-z=0-hm}
\end{eqnarray}

Eq. (\ref{results-s-wave-z=0-hm}) reveals another
main conclusion of this paper:
The boundary-transport-contributions
are absent in the expressions for the non-equilibrium magnetization,
$m$, and the induced spin-dependent boundary voltage, $V_s$,
when the ferromagnet is half-metallic. For $m$, this 
is the result of the simple relationship between the
spin current and total current (Eq. (\ref{I-spin-s-wave-z=0-hm})),
which arises whenever the Andreev reflection is absent.
(Note that the total current $I$ is fixed.)
For $V_s$, the reasoning leading to this conclusion is slightly
more subtle. When the Andreev reflection is absent,
the boundary-transport-contributions give rise to exactly
the same prefactors in the two currents in 
Eq. (\ref{I-sigma-closed-s-wave-z=0-open}). Since the
boundary voltage $V_{\sigma}$ is determined by balancing these
two terms, the boundary-transport factors cancel out exactly
in the expression for $V_s$.

The above cancellation argument does not
apply to the induced current $I_{\sigma}$.
Indeed, $I_s$ does contain the boundary-transport factors.

{\em $s-$wave, large barrier:~~}
Consider now the limit of a large barrier. 
The results for the pair, single-particle, and
spin currents parallel Eq. (\ref{I-s-wave-z=0}),
except that the factors $P_{s,0}$, $S_{s,0}$, $M_{s,0}$,
and $Y_{s,0}$ are replaced by
\begin{eqnarray}
P_{s,\infty} &&= (1/z^4) g_0 \alpha_{pair}
t_{\uparrow} t_{\downarrow} \nonumber\\
S_{s,\infty} &&= \sqrt{\pi/2}(1/z^2) g_0 \alpha_{sp} 
(t_{\uparrow}
+t_{\downarrow}) \nonumber\\
M_{s,\infty} &&= \sqrt{\pi/2}(1/z^2) g_0 \alpha_{sp} (t_{\uparrow}
-t_{\downarrow}) \nonumber\\
Y_{s,\infty}(T)  &&= 
\sqrt{\Delta / {k_BT} } e ^{-\Delta / k_BT}
\label{results-s-wave-z=large}
\end{eqnarray}
respectively.
The resulting expressions for $m$, $I_s$ and $V_s$ are also
given by Eqs. 
(\ref{m-s-wave-z=0}, \ref{I-s-s-wave-z=0}, \ref{V-s-s-wave-z=0}), 
with an appropriate substitution by the
factors given in Eq. (\ref{results-s-wave-z=large}).

The pair current is of order $1/z^4$. Both the single-particle
current and spin current, on the other hand,
are of order $1/z^2$. The latter is to be expected, since 
in the large barrier limit the single-particle transport
can be described in terms of a tunneling picture. 
For the same reason, the temperature dependence of $I_{sp}$ and
$I_{spin}$ in this case should reflect simply the thermal smearing
of the single-particle density of states in the superconductor.
Indeed, $Y_{s,\infty}(T)$ corresponds to the low temperature limit
of the Yosida function.

Whenever the Andreev reflection is non-negligible,
each of the three quantities, $m$, $V_s$ and $I_s$,
again depends on a different combination of the boundary transport 
and bulk transport properties.

When the potential barrier is so strong that the temperature-independent
$1/z^4$ contribution is negligible compared to the temperature-dependent
$1/z^2$ terms, the Andreev process is absent.
Here again, while the boundary transport terms still appear in $I_s$,
they are cancelled out in $m$ and $V_s$.

{\em $d-$wave case:~~}
We now turn to the case of a $d_{x^2-y^2}$ superconductor. Here
we will focus on the case when both the FM1-S and S-FM2
interfaces involve the $\{110\}$ surface of the cuprates.
Consider first the limit of a vanishing barrier.
The result for the currents again parallels
Eq. (\ref{I-s-wave-z=0}), 
with the prefactors replaced, respectively, by
\begin{eqnarray}
P_{d,0} &&= 16 g_0 \alpha_{pair} 
t_{\uparrow} t_{\downarrow}/(t_{\uparrow} + t_{\downarrow})^2 
\nonumber\\
S_{d,0} &&= (\ln 2 ) g_0 \alpha_{sp} 
[C_{\uparrow} t_{\uparrow}+ (\uparrow \rightarrow \downarrow)]/
(t_{\uparrow} t_{\downarrow}+1)^2 \nonumber\\
M_{d,0} &&= (\ln 2 ) g_0 \alpha_{spin} 
[C_{\uparrow} t_{\uparrow}
- (\uparrow \rightarrow \downarrow)]/
(t_{\uparrow} t_{\downarrow}+1)^2 \nonumber\\
Y_{d,0}(T)  &&= k_B T/\Delta
\label{results-d-wave-z=0}
\end{eqnarray}
where 
$C_{\uparrow} = (t_{\downarrow} + 1)^2
+ 4 (3-\pi) t_{\downarrow}
-(19/3 - 2 \pi ) (t_{\downarrow} -1 )^2$.

For the large barrier limit, 
we found $P_{d,\infty} = P_{d,0}$,
$Y_{d,\infty}(T)  = Y_{d,0}(T)$, and 
\begin{eqnarray}
S_{d,\infty} &&= (4-\pi)(\ln 2 / 2) (1/z^2) \alpha_{sp} 
(t_{\uparrow}+t_{\downarrow}) \nonumber\\
M_{d,\infty} &&= (4-\pi)(\ln 2 / 2) (1/z^2) \alpha_{spin} 
(t_{\uparrow}-t_{\downarrow})
\label{results-d-wave-z=large}
\end{eqnarray}
Note that, while $I_{sp}$ and $I_{spin}$ are still of
order $1/z^2$, $I_{pair}$ is of order $1/z^0$. The latter 
reflects the formation of the
Andreev-bound states\cite{Hu,Kashiwaya,Ting,Sauls}.

To summarize, we have studied the effect of spin injection into $s-$ and
$d-$wave superconductors. Through the explicit results in the small and
larger barrier limits, we conclude that Andreev-reflection makes
the different physical properties measured in spin injection experiments
reflect very different combinations of the boundary and bulk spin
transport properties.
For the purpose of isolating bulk spin transport properties,
it is desirable to make the Andreev contribution as small as
possible. In this case, the non-equilibrium magnetization
and the spin-dependent boundary voltage depends solely on the
bulk spin transport properties. This occurs in two limits.
One is the limit of a strong barrier in tunneling geometries such that
the Andreev bound states are absent.  The other is
for the half-metallic ferromagnet. 

This work has been supported in part by the NSF Grant No. DMR-9712626,
a Robert A. Welch grant, and an A. P. Sloan Fellowship. 
While this paper was under preparation we learned of a
preprint\cite{Kashiwaya} of Kashiwaya et al. who calculated the 
non-linear spin current for the case of a $d-$wave superconductor.


\begin{references}
\bibitem{magnon} The spin waves have a super-ohmic spectral
function and are hence unimportant for our analysis.
\end{references}
\end{document}